\documentclass[12pt]{article}
\usepackage{graphics}
\begin{document}
\title{From turbulence to financial time series}
\author{\normalsize B. Holdom\\\small {\em Department of Physics, University
of Toronto}\\\small {\em Toronto, Ontario,} M5S1A7,
CANADA\\\footnotesize holdom@utcc.utoronto.ca}\date{}\maketitle
\begin{abstract}
We develop a framework especially suited to the autocorrelation
properties observed in financial times series, by borrowing from the
physical picture of turbulence. The success of our approach as applied
to high frequency foreign exchange data is demonstrated by the overlap
of the curves in Figure (1), since we are able to provide an
analytical derivation of the relative sizes of the quantities
depicted. These quantities include departures from Gaussian
probability density functions and various two and three-point
autocorrelation functions.\\47.27.Eq, 89.90.+n, keywords: time series, turbulence, autocorrelation
\end{abstract}
\baselineskip 19pt

\baselineskip 25pt

Financial time series and the unexpected scaling behaviors they
display have recently attracted the attention
\cite{a1}--\cite{a9} of the theoretical
physics community, as large data sets have become available. From a
financial times series \({\it x}(t)\) and the corresponding set of log price changes \(\Delta x\equiv \ln({\it x}(t + \Delta t)) - \ln({\it x}(t))\) for time delay \(\Delta t\), one constructs the probability density functions (PDFs) \({P_{\Delta t}}(\Delta x)\). There are three basic observations \cite{a6}. 1) \({P_{\Delta t}}(\Delta x)\) often departs in a leptokurtic manner (sharper peak and close to
exponential tails) from the Gaussian shape expected for a random walk.
2) Scale invariance is broken in the sense that the shape of \({P_{\Delta t}}(\Delta x)\) evolves and gradually becomes more similar to a Gaussian as \(\Delta t\) increases. More striking is that this approach to a Gaussian is
slower than expected if the individual price changes were
independently and identically distributed with finite variance. This
anomalous scaling indicates that there is some kind of correlation
between the price changes at different times (autocorrelation). 3)
Although there is negligible autocorrelation observed in the log price
changes themselves, there is significant autocorrelation observed in
the various functions of the log price changes. It is seen for example
in the square of the log price changes, and thus is referred to as
volatility clustering.

We make the following definitions, where we assume that the PDFs have
mean \(\mu =0\) and variance \(\sigma ^{2}=1\) (otherwise replace \(\Delta x\) by \((\Delta x - \mu )/\sigma \), etc.).
\begin{eqnarray}&&{\it F}({\it f}(x), \Delta t)=\frac {\langle {\it f}(
\Delta x)\rangle }{\langle {\it f}(\Delta x){\rangle _{{\rm Gauss}}}} - 1{\label{e5}}\\&&{\it G}({\it f}(x), \tau )=\frac {\langle {\it f}(\Delta x
){\it f}(\Delta y)\rangle  - \langle {\it f}(\Delta x)
\rangle ^{2}}{\langle ({\it f}(\Delta x) - \langle {\it f
}(\Delta x)\rangle )^{2}\rangle }{\label{e6}}\\&&{\it H}({\it f}(x), \tau )=\frac {\langle {\it f}(\Delta x
){\it f}(\Delta y){\it f}(\Delta z)\rangle  - 2\langle {\it f}(\Delta x){\it f}(\Delta z)\rangle \langle 
{\it f}(\Delta x)\rangle  + \langle {\it f}(\Delta x)
\rangle ^{3}}{\langle ({\it f}(\Delta x) - \langle {\it f}(
\Delta x)\rangle )^{2}\rangle ^{{\frac {3}{2}}}}{\label{e7}}\end{eqnarray}
 \(F\) for various functions \({\it f}(x)\) will provide different measures of the departures from a Gaussian
PDF, as a function of the time delay \(\Delta t\). \(G\) is the autocorrelation function for a lag time \(\tau \) where \(\Delta x\) is as above and \(\Delta y\equiv \ln({\it x}(t - \tau  + \Delta t))
 - \ln({\it x}(t - \tau ))\).  \(H\) is a three-point autocorrelation function where \(\Delta z\equiv \ln({\it x}(t - \frac {\tau }{2} + 
\Delta t)) - \ln({\it x}(t - \frac {\tau }{2}))\), and it is constructed so that it receives no contributions from
autocorrelations on scales smaller than \(\tau \). Note that for the autocorrelations, \(\Delta t\) is typically constrained to be much less than \(\tau \). \(F\), \(G\) and \(H\) all vanish for a Gaussian random walk.

We consider the data set made freely available in association with the
Santa Fe Competition \cite{d}. It consists of 329112
quotes (bid and ask) of the Swiss franc--US dollar exchange rate
between May 20, 1985 to April 12, 1991. The time between price quotes
is on average a few minutes, but it is very irregular. From this data
we extracted a time series with a fixed time step of 10 minutes and a
total of 84914 prices. To do this we averaged the bid and ask quotes,
used linear interpolation to determine a price at each time, and
eliminated the prices changes between days.

In Fig.\ (1) we plot as functions of \(\Delta t\) or \(\tau \) the quantities  \({\it F}({f_{a}}(x), \Delta t)/{\zeta _{a}}\), \({\it G}({f_{a}}(x), \tau )/{\xi _{a}}\) and \({\it H}({f_{a}}(x), \tau )/{\kappa _{a}}\) for \(a=1, 2, {...}, 6\). The functions \({f_{a}}(x)\) and our calculated constants \(({\zeta _{a}}, {\xi _{a}}, {\kappa _{a}})\) (which are independent of the normalization of \({f_{a}}(x)\)) are specified as follows, where \(r\equiv  \left|   x   \right| \).
\begin{equation}
\begin{array}{|c|c|c|c|c|c|c|c|c|c|}\hline
{f_{a}}(x) & r^{1/4} & \sqrt{r} & {\displaystyle \frac {1}{1
 + r}}  & {\displaystyle \frac {1}{3 + r}}  & {\displaystyle 
\frac {1}{1 + r^{2}}}  & {\displaystyle \frac {1}{(1 + r)^{2}}} 
 \\\hline
{\zeta _{a}} & -.0574 & -.0944 & .0772 & .0315 & .0961 & .159
 \\\hline
{\xi _{a}} & .286 & .310 & .299 & .315 & .309 & .270 \\\hline
{\kappa _{a}} & 1.026 & .617 & 1.045 & 2.129 & .815 & .586\\\hline
\end{array}{\label{e10}}
\end{equation}
The striking result is that the quantities plotted in Fig. (1) are
nearly independent of the function \({f_{a}}(x)\). This paper will provide an understanding of this result through the
derivation of the \(({\zeta _{a}}, {\xi _{a}}, {\kappa _{a}})\) constants. From the range of these constants we see that the
functions \({f_{a}}(x)\) are probing the characteristics of the price changes in different
ways. Our method also applies to any other functions of choice, and at
the end we shall comment on functions which are more sensitive to very
large price changes.

We will motivate a one parameter family of PDFs with the following
properties. 1) They are derived in closed form. 2) They are
leptokurtic with exponential tails. 3) They are closed under
convolution. 4) Their derivation is linked to autocorrelation.

The similarity between PDFs in finance and turbulence has recently
been emphasized in \cite{a2}. For turbulence in a fluid,
PDFs are obtained for the velocity differences at two points separated
by some distance. The leptokurtic shapes of these PDFs is a reflection
of the intermittency of turbulent flow \cite{f}. The
origin of this intermittency lies in the coming and going of coherent
structures of various sizes, or eddies, in the velocity field. The
essential point is that volatility in the velocity differences can be
correlated over distance scales typical of the sizes of the eddies. It
is the idea of `volatility structures' of varying time scales, rather
than size, which we will carry over to financial time series, where
price changes replace velocity differences. We will discuss further
the analogy between turbulence and financial time series at the end.

To develop a simplified picture we define a volatility structure (VS)
of time scale \(T\) as follows. Take the total time of the series and split it up into
intervals of length \(T\). Let the volatility be constant on each interval, but allow it to
vary from one time interval to the next. The distribution of these
values defines a PDF for volatilities on scale \(T\). Our goal is to account for the cumulative effects of VSs occurring
on all time scales.

First we consider the effect of a single VS of time scale \(T\) on the PDF \({P_{\Delta t}}(\Delta x)\) for \(T\gg \Delta t\). We consider all pairs of prices at times differing by \(\Delta t\); over each of these \(\Delta t\) intervals the volatility will be constant (ignoring those few
intervals which happen to straddle a boundary), but the value of this
volatility can be different for widely separated intervals. Thus the
PDF will follow by summing over all random walks of duration \(\Delta t\) for all values of the constant volatility. This leads to a
superposition of Gaussian PDFs,
\begin{equation}{{\cal{P}}_{\eta }}(\sigma , \Delta x)=\int _{0}^{\infty }{p_{\eta }}(S){P_{{\rm Gauss}}}(S\sigma
 , \Delta x)dS{,\label{e1}}\end{equation}
where \({P_{{\rm Gauss}}}(\sigma , \Delta x)=\frac {1}{\sqrt{2\pi }\sigma }e^{{ - \frac {(\Delta x)^{2}}{2\sigma ^{2}}}}\) and \(\sigma ^{2}\) is the variance.

 \({p_{\eta }}(S)\) with \(S\geq 0\) is the PDF for the volatility. To produce a family of PDFs with the
properties listed above we make the following
choice.\footnote{An alternative choice for \({p_{\eta }}(S)\), proposed in both the financial \cite{g} and turbulence
\cite{h} literature and advocated more recently
\cite{a2,a8}, is the lognormal PDF. The drawback is that
the resulting family of PDFs cannot be obtained in closed form, and
they are not closed under convolution in the sense we describe.}
\begin{equation}{p_{\eta }}(S)=\frac {2\eta ^{\eta }}{\Gamma (\eta )}S^{{2\eta  - 1}}e^{{ - \eta S^{2}}}{\label{e2}}\end{equation}
We note that this includes the simple Gaussian, \({p_{\frac {1}{2}}}(S)=\sqrt{\frac {2}{\pi }}e^{{ - \frac {S^{2
}}{2}}}\). The volatility PDF has a mean which approaches unity from below as \(\eta \) increases, while its variance about this mean approaches
\begin{equation}{\sigma _{{\rm vol}}^2}\approx \frac {1}{4\eta }{.\label{e4}}\end{equation}
Thus for large \(\eta \), \({p_{\eta }}(S)\) becomes sharply peaked around \(S=1\) and \({P_{\eta }}\) tends towards \({P_{{\rm Gauss}}}\).

The integral in (\ref{e1}) can be evaluated to yield \({{\cal{P}}_{\eta }}(\sigma , \Delta x)={{\cal{Q}}_{1, \eta
 }}(\sigma , r)\) where \(r= \left|   \Delta x   \right| \), \({{\cal{Q}}_{i, \eta }}(\sigma , r)={{\cal{Q}}_{i, \eta }}(
1, r/\sigma )/\sigma \) and
\begin{equation}{{\cal{Q}}_{i, \eta }}(1, r)=\frac {1}{\pi ^{{\frac {i}{2}}}
\Gamma (\eta )}2^{{1 - \frac {i}{4} - \frac {\eta }{2}}}\eta ^{{\frac {i}{4}
 + \frac {\eta }{2}}}r^{{\eta  - \frac {i}{2}}}{K_{\eta  - \frac {i}{2}}}(\sqrt{2\eta }r){.
\label{e11}}\end{equation}
 \({K_{\nu }}(x)\) is the modified Bessel function of the second kind and \(i\neq 1\) will be of interest below. These PDFs have a distinctive leptokurtic
shape which becomes more prominent for decreasing \(\eta \). For any positive integer \(n\), \({{\cal{P}}_{n}}(1, \Delta x)\)  is simply a polynomial in \( \left|   \Delta x   \right| \) of degree \(n - 1\) times the exponential \(e^{{ - \sqrt{2n} \left|   \Delta x   \right| }}\). The appearance of these exponential tails can be traced to the
Gaussian factor in (\ref{e2}). The simplest member of
this family of PDFs is purely exponential, \({{\cal{P}}_{1}}(1, \Delta x)=\frac {1}{\sqrt{2}}e^{{ - \sqrt{2} \left|   \Delta x   \right| }}\).

We now consider the PDF \({P_{\Delta t}}(\Delta x)\) which results from a single VS with \(T<\Delta t\). For example with \(\Delta t=nT\) we must consider the convolution
\begin{equation}\int d{x_{1}}{...}{{\it dx}_{n - 1}}{{\cal{P}}_{\eta }}(\sigma , {x_{n}} - {x_{n - 1}}){{\cal{P}
}_{\eta }}(\sigma , {x_{n - 1}} - {x_{n - 2}}){...}{{\cal{P}}_{\eta }}(\sigma , {x_{1}} - {x_{0}}){\label{e3}}\end{equation}
where the \({x_{i}}\) are the log prices at regularly spaced intervals \({t_{i}} - {t_{i - 1}}=T\) . This convolution accounts for the variable volatility on each
interval. We now make use of the fact that our family of PDFs is
closed under convolution. In particular the results of
\cite{i} are sufficient to prove the following relation
for half-integer \({\eta _{1}}\) and \({\eta _{2}}\).
\begin{equation}\int dy{{\cal{P}}_{{\eta _{1}}}}(\sqrt{{\eta _{1}}}\sigma , x - y)
{{\cal{P}}_{{\eta _{2}}}}(\sqrt{{\eta _{2}}}\sigma , y - z)
={{\cal{P}}_{{\eta _{1}} + {\eta _{2}}}}(\sqrt{{\eta _{1}} + {\eta _{2}}}\sigma , x - z)\end{equation}
Thus the PDF \({P_{\Delta t}}(\Delta x)\) determined by (\ref{e3}) is \({{\cal{P}}_{n\eta }}(\sqrt{n}\sigma , {x_{n}} - {x_{0}})\).

There are three points worth stressing for this case of a VS with \(T<\Delta t\). 1) The variance of \({P_{\Delta t}}(\Delta x)\) increases exactly linearly with \(\Delta t\) for fixed \(T\). (This is trivially true for the previous case, \(T\gg \Delta t\).) This linear growth of variance is a well documented feature of
financial time series \cite{a6}, which can be related to
the vanishing of \({\it G}({\it f}(x), \tau )\) for \({\it f}(x)=x\) \cite{a5}. 2) From the relation (\ref{e4})
between \({\sigma _{{\rm vol}}^2}\) and \(\eta \)  we see that the effective variance in volatilities \({\sigma _{{\rm vol}}^2}\) on time scale \(\Delta t\) is decreasing as \(1/\Delta t\). In other words as the number \(n\) of intervals with independently fluctuating volatilities increases
the effective PDF tends towards a Gaussian, in a manner as expected
from the central limit theorem \cite{c}. 3) We now see
that the whole family of PDFs in (\ref{e1}) are nothing
but convolutions of the simple case when the volatility PDF \({p_{\eta }}(S)\) is a Gaussian (\(\eta =1/2\)).

We next consider the autocorrelation between price changes \(\Delta x\) and \(\Delta y\), each occurring over a time step \(\Delta t\), but separated by a lag \(\tau >\Delta t\). For a VS with \(T\gg \tau \) we must sum over all pairs of random walks of duration \(\Delta t\) for all values of a common volatility. The joint PDF is then
\begin{equation}{{\cal{J}}_{\eta }}(\sigma , \Delta x, \Delta y)=\int _{0}^{\infty }{p_{\eta }}(S){P_{{\rm Gauss}}}(S\sigma
 , \Delta x){P_{{\rm Gauss}}}(S\sigma , \Delta y)
dS.\end{equation}
More generally an \(i\)-point joint PDF can be defined in the corresponding way, and we
obtain
\begin{equation}{{\cal{J}}_{\eta }}(\sigma , \Delta {x_{1}}, \Delta {x_{2}}, {...}, \Delta {x_{i}})={{\cal{Q}}_{i, \eta }}(
\sigma , \sqrt{(\Delta {x_{1}})^{2} + (\Delta {x_{2}})^{2}
 + {...} + (\Delta {x_{i}})^{2}}),\end{equation}
where \({{\cal{Q}}_{i, \eta }}(1, r)\) is defined in (\ref{e11}).

From this joint PDF we may extract for example the PDF for the
following combination of the \(m\) adjacent \(\Delta x\)'s within time \(m\Delta t\): \({v_{m}}\equiv \sqrt{\sum _{{\rm adjacent}}(\Delta x)
^{2}}\).\footnote{A very similar quantity was recently studied
in \cite{a9}.} Thus for VSs with \(T\gg m\Delta t\) we obtain the PDF \({{\cal{R}}_{m, \eta }}({v_{m}})=2\pi ^{{\frac {m}{2}}}{v_{m}}^{{m - 1}}{{\cal{Q}}_{m, \eta }}(\sigma , {v_{m}})/
\Gamma (\frac {m}{2})\). This cannot be directly compared with data because it does not
account for VSs with \(T\ \lower 2pt \hbox{$\buildrel<\over{\scriptstyle{\sim}}$}\ m\Delta t\).

We return to the quantities \(F\), \(G\), and \(H\) in (\ref{e5}--\ref{e7}), for which we can
account for VSs occurring on all time scales \(T\). For \(G\) and \(H\) only VSs with \(T\) larger than \(\tau \) contribute, and the cumulative effect of such VSs can be represented
by an effective \({\sigma _{{\rm vol}}^2}({}\tau )\). From the inverse relation (\ref{e4}) between \({\sigma _{{\rm vol}}^2}\) and \(\eta \) this corresponds to an effective \(\eta \) which will specify \({{\cal{J}}_{\eta }}(\sigma , \Delta x, \Delta y, {...})\), which in turn enters the determination of \(G\) and \(H\). \(G\) and \(H\) vanish only in the large \(\eta \) limit.

For the PDF \({P_{\Delta t}}(\Delta x)\), VSs of time scales both larger and smaller than \(\Delta t\) contribute. We have seen how each VS with \(T<\Delta t\) makes a contribution to an effective \({\sigma _{{\rm vol}}^2}({}\Delta t)\) suppressed by \(T/\Delta t\). These contributions can be added to the contribution of VSs larger
than \(\Delta t\) to yield a total effective \({\sigma _{{\rm vol}}^2}({}\Delta t)\). The corresponding effective \(\eta \) then specifies a PDF \({{\cal{P}}_{\eta }}(\sigma , \Delta x)\) which can be used to determine \(F\). Unlike the case of a single VS with \(T<\Delta t\), the effective \({\sigma _{{\rm vol}}^2}({}\Delta t)\) may now fall slower than \(1/\Delta t\). This corresponds to the slow approach to a Gaussian, which is the
anomalous scaling observed in the data. Thus in our picture it is the
presence of VSs on all time scales which underlies the link
\cite{a6,a5} between this anomalous scaling and the
nonvanishing autocorrelations.

To compare with data we first use the 1, 2, and 3-point PDFs, \({{\cal{P}}_{\eta }}\) and \({{\cal{J}}_{\eta }}\), to determine \(F\), \(G\), and \(H\) as functions of \(\eta \). 
\begin{eqnarray}&&{\it F}({f_{a}}(x), \eta )=\frac {{\zeta _{a}}(\eta )}{\eta }{,\ \ \ \ \ }{\it G}({f_{a}}(x), \eta )=\frac {{\xi _{a}
}(\eta )}{\eta }{,\ \ \ \ \ }{\it H}({f_{a}}(x), \eta )=\frac {{\kappa 
_{a}}(\eta )}{\eta }{\label{e20}}\end{eqnarray}
We find that \({\zeta _{a}}(\eta )\), \({\xi _{a}}(\eta )\) and \({\kappa _{a}}(\eta )\) are rather slowly varying functions which tend to constants for
large \(\eta \). In principle we could extract \({\it F}({f_{a}}(x), \Delta t)\), \({\it G}({f_{a}}(x), \tau )\), and \({\it H}({f_{a}}(x), \tau )\) from the data and by comparing with (\ref{e20}),
determine the appropriate \(\eta \) for each \(\Delta t\) and \(\tau \) respectively. The prediction would be that the values of \(\eta \) and the values of \(F/\zeta \), \(G/\xi \) and \(H/\kappa \) are independent of \({f_{a}}(x)\). In fact for the range of \(\eta \) we are considering we can, to good approximation, replace the
functions \({\zeta _{a}}(\eta )\), \({\xi _{a}}(\eta )\) and \({\kappa _{a}}(\eta )\) by constants. The values listed in  (\ref{e10})
correspond to (\({\zeta _{a}}(3/2)\), \({\xi _{a}}(3/2)\), \({\kappa _{a}}(3/2)\)), which in turn yield the results in Fig.\ (1). In
addition to the overlap of the curves we also note the close
similarity of Figs.\ (1B) and (1C). This strongly supports
our common description of all \(i\)-point autocorrelation functions in terms of the function \({{\cal{Q}}_{i, \eta }}(\sigma , r)\) in (\ref{e11}).

In general our model provides a good description of the PDFs and the
autocorrelations, with the exception being quantities sensitive to
abnormally large price changes. For example when \(F/\zeta \) with \({\it f}(x)=x^{4}\) is extracted from the data we obtain the top curve in
Fig.\ (1A). (\({\it F}(x^{4}, \eta )\) is directly related to the kurtosis of our PDFs, and in this case we
have simply \(\zeta (\eta )=1\).) This indicates that the extreme tails of the true PDFs fall off
somewhat slower than expected, especially for shorter time delays.
More striking are the autocorrelations involving large price changes;
for example \(G/\xi \) and \(H/\kappa \) with \({\it f}(x)=x^{2}\) gives the bottom curves in Figs.\ (1B) and (1C)
respectively. Thus large price changes show much less autocorrelation
than expected. This suggests that large price changes more strongly
reflect the reaction of the markets to external events.  

We conclude by commenting on the analogy with turbulence in a fluid
\cite{a2,a4,a6,a3,a8}. The PDFs for turbulence evolve as
a function of the distance \(r\) separating the two points at which the velocities are measured. We
have presented a model for this evolution based on a formulation
related to the present work \cite{i}. A basic observation
is that the evolution becomes more transparent when expressed in terms
of \(\tau {\ \propto\ }r^{{2/3}}\). (In fact \(\tau \) is a physical time scale in turbulence, the so-called turnover time
\cite{f}.) In particular the variance of the PDF is
proportional to \(\tau \), while its shape evolves from the leptokurtic towards the Gaussian
as \(\tau \) increases, making the analogy between financial time series and
turbulence very close. The difference for turbulence in a fluid is
that the data is quite consistent \cite{i} with a speed
of approach to the Gaussian as described by the central limit theorem,
\({\sigma _{{\rm vol}}^2}{\ \propto\ }1/
\tau \). Thus contrary to financial time series, volatility structures
(eddies in this case) on large \(\tau \) scales are not sufficiently strong to cause a significant slowing
down in the approach to the Gaussian. The turbulence in the financial
markets is stronger in this sense than the turbulence observed in a
fluid.

\vspace{5ex}
\section*{Acknowledgement}
This research was supported in part by the Natural Sciences and
Engineering Research Council of Canada.

\vspace{10ex}
\noindent Figure (1): A, B, and C each display six curves
with significant overlap. \({\it F}({f_{a}}(x), \Delta t)/{\zeta _{a}}\), \({\it G}({f_{a}}(x), \Delta t)/{\xi _{a}}\) and \({\it H}({f_{a}}(x), \Delta t)/{\kappa _{a}}\) are defined in (\ref{e5}--\ref{e7}) with
the six functions \({f_{a}}(x)\) and constants (\({\zeta _{a}}\), \({\xi _{a}}\), \({\kappa _{a}}\)) specified in (\ref{e10}). The additional dashed line
in A corresponds to \({\it f}(x)=x^{4}\), while the dashed line in each of B and C corresponds to \({\it f}(x)=x^{2}\). To produce B and C we have removed from the data the well known
correlation between volatility and time of day. This changes the shape
of the curves but not their overlap.

\begin{center} \includegraphics{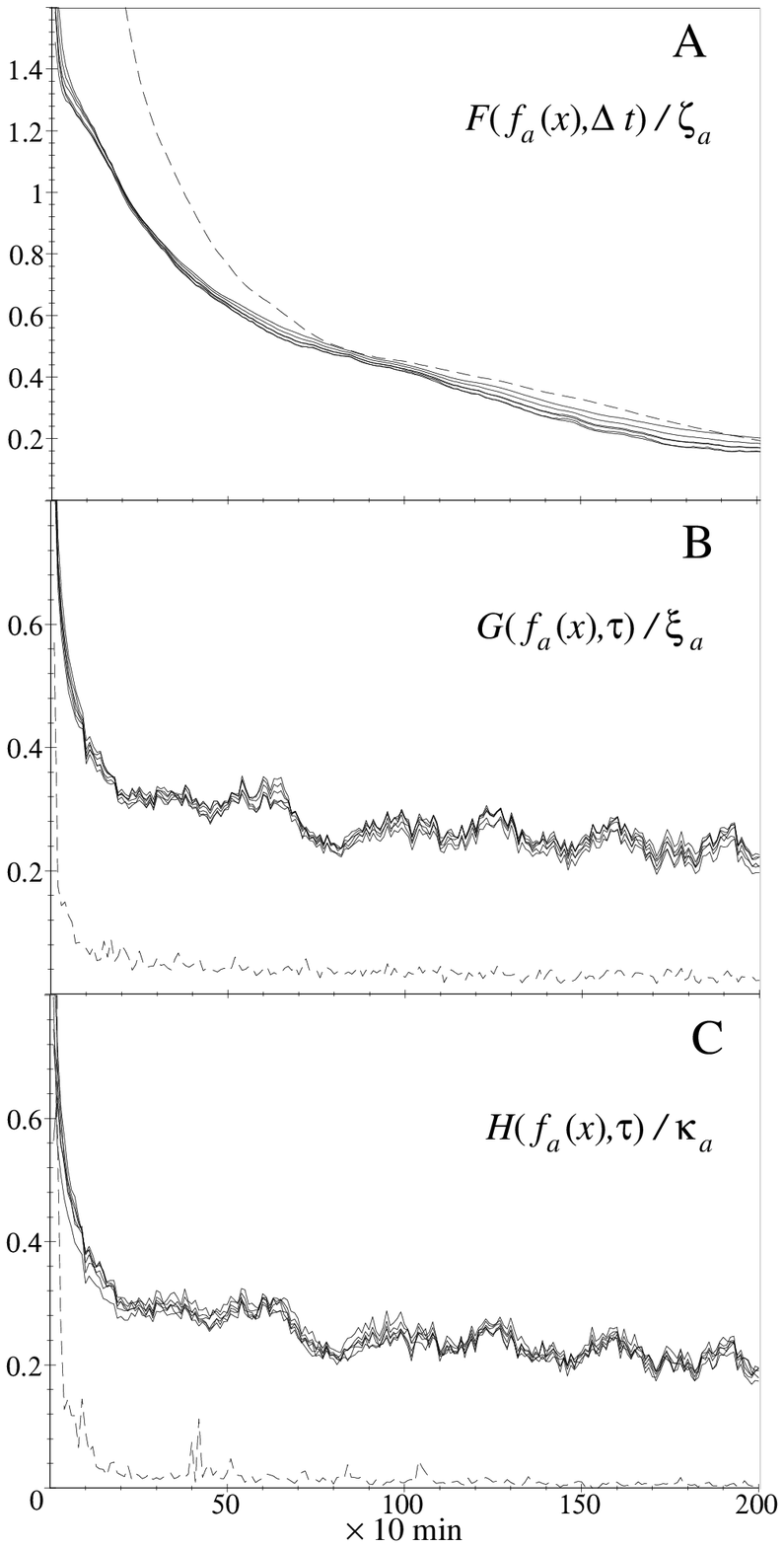}
\end{center}

\end{document}